\documentclass[showpacs,showkeys,11pt,
preprint,preprintnumbers,nofootinbib,
groupedaddress,superscriptaddress,amsmath,amssymb]{revtex4}
\usepackage{amsfonts}
\usepackage{graphics}
\usepackage{epsfig}
\usepackage{url}
\usepackage{multirow}
\usepackage{feynmp}
\newcommand {\be}{\begin{equation}}
\newcommand {\ee}{\end{equation}}
\newcommand {\ba}{\begin{eqnarray}}
\newcommand {\ea}{\end{eqnarray}}

\newcommand {\tanb}{$\tan\beta~$}
\newcommand {\ra}{\rightarrow}
\newcommand {\sinb}{s_{\beta}}
\newcommand {\cosb}{c_{\beta}}
\newcommand {\sbma}{s_{\beta-\alpha}}
\newcommand {\cbma}{c_{\beta-\alpha}}
\newcommand {\sw}{s_{W}}
\newcommand {\tb}{t_{\beta}}
\newcommand {\ctb}{t_{\beta}^{-1}}

\begin{document}
\title{Observability of Charged Higgs through Triple or Double Higgs Production in a General Two Higgs Doublet Model at an $e^+e^-$ Linear Collider}
\author{M. Hashemi}
\email{hashemi_mj@shirazu.ac.ir}
\affiliation{Physics Department and Biruni Observatory, College of Sciences, Shiraz University, Shiraz 71454, Iran}
\author{I. Ahmed}
\email{ijaz.ahmed@cern.ch}
\affiliation{COMSATS Institute of Information Technology (CIIT), Islamabad 44000, Pakistan}
\affiliation{National Center for Particle Physics, Institute of Research Management \& Monitoring, University of Malaya, 50603 Kuala Lumpur, Malaysia}

\begin{abstract}
In this paper the charged Higgs signal through triple or double Higgs production in a general two Higgs doublet model (2HDM) is studied. The main production process is $e^+e^- \ra H^+H^-H^0$ followed by the charged Higgs decay to a pair of $\tau \nu$ and the neutral Higgs decay to $b\bar{b}$. The alternative process $H^+W^-H^0$ is also included as a source of charged Higgs signal in the analysis. The focus is on a future $e^+e^-$ linear collider operating at $\sqrt{s}=1.5$ TeV. The final state under consideration ($\tau^+ \tau^- b \bar{b} E^{\textnormal{miss}}_{T}$) is suitable for electroweak background rejection using the $b-$tagging tools. It is shown that although the signal cross section is small, with a reasonable background suppression, high signal significance values are achievable at an integrated luminosity $500~fb^{-1}$ depending on the charged Higgs mass, \tanb and the CP-odd neutral Higgs mass. Finally results are quoted in terms of the signal significance for charged Higgs in the mass range $170~<~m_{H^{\pm}}~<~400$ GeV. 
\end{abstract}

\maketitle

\section{Introduction}
\begin{figure}[h]
 \begin{center}
 \includegraphics[width=0.6\textwidth]{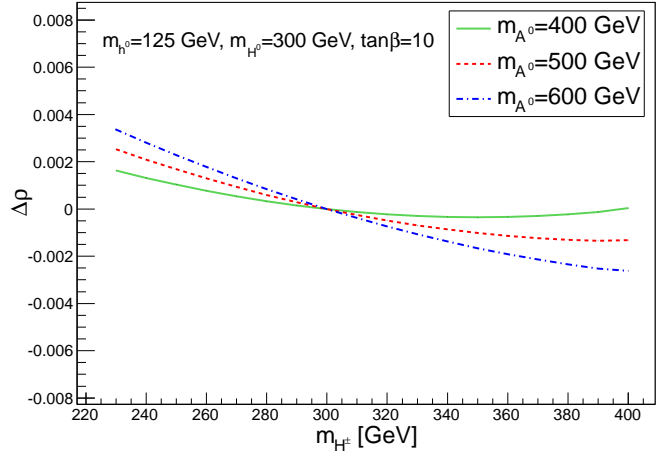}
 \end{center}
 \caption{$\Delta \rho$ parameter as a function of charged Higgs mass for different $m_{A}$ values.}
 \label{drho}
 \end{figure}

The missing element of the Standard Model of Particle Physics has already been observed by the two LHC experiment detectors, i.e., ATLAS and CMS \cite{125atlas,125cms}. The mass of the particle is near 125 GeV. The question of possibility of existence of further Higgs bosons is, however, still open. There are certainly motivating theoretical arguments for SM extentions; One of those being the Higgs boson mass quadratic divergence when radiative corrections are included. A natural solution to this problem is Supersymmetry \cite{susy,susy2} which requires a non-minimal Higgs sector. In order to build up a SUSY model, at least two Higgs doublets are required.  The Minimal Supersymmetric Standard Model (MSSM) is the simplest example of a supersymmetric model which belongs to two Higgs doublet models (2HDM) \cite{Martin}. In a general 2HDM, the Higgs sector consists of two charged Higgs bosons, $H^{\pm}$, two CP-even neutral Higgs bosons, $h^0,~H^0$, and a CP-odd neutral Higgs, $A^0$. The lightest neutral Higgs boson, $h^0$, is taken to be SM-like and is the candidate for the signal observed at LHC. 
The study of two Higgs doublet models is therefore necessary as the first step to construct a model whose structure is a little more complicated than SM.\\  
There have been searches for Higgs bosons of non-SM type at LEP and Tevatron. A charged Higgs with $m_{H^{\pm}}<89~ \textnormal{GeV}$ has been excluded by LEP for all \tanb values \cite{lepexclusion1}. The Tevatron searches by D0 \cite{d01,d02,d03,d04} and CDF \cite{cdf1,cdf2,cdf3,cdf4} restrict the charged Higgs mass to be in the range $m(H^{\pm}) > 80 $ GeV for $2 <$ tan$\beta < 30$. The current published results from LHC exclude \tanb $>10$ with a charged Higgs boson with $m_{H^{\pm}}=90$ GeV \cite{atlasdirect,cmsdirect}. The above results are followed by the indirect limit from $b \ra s\gamma$ studies by the CLEO collaboration which exclude a charged Higgs mass below 300 GeV at 95 $\%$ C.L. in 2HDM Type II with \tanb$>2$ \cite{B1}. In general in terms of 2HDM types the current conclusion is $m_{H^{\pm}}~>~300$ GeV in 2HDM Type II and III. However, this lower bound does not apply to types I and IV at high \tanb \cite{main2hdm}. In addition to the above constraints, there are recent results from the CMS and ATLAS collaborations based on direct and indirect searches for the charged Higgs boson within the MSSM framework. The CMS collaboration restricts a neutral MSSM Higgs boson to be heavier than 200 GeV with \tanb = 10, assuming a light neutral SM Higgs boson around 125 GeV \cite{h2tautauCMS}. Therefore, an additional neutral Higgs should be heavier than at least 200 GeV with \tanb less than 10. This limit, does not exclude heavier Higgs bosons. Based on this observation, and to allow for high \tanb values, a neutral Higgs boson with a mass of 300 GeV is assumed in this paper. New results from ATLAS based on direct searches for the charged Higgs indicate no charged Higgs lighter than 160 GeV with \tanb$>20$ \cite{atlconf2013}. Comparing these results with CMS search report \cite{cmschnew} and the fact that there is no exclusion for a charged Higgs boson heavier than 170 GeV by CMS, the starting point for the charged Higgs mass in this paper is taken as 170 GeV. \\
The chosen mass points are also consistent with $B_s \rightarrow \mu^+\mu^-$ studies. The present measurement accuracy of BR($B_s \rightarrow \mu^+\mu^-$) at LHCb indicates that including results from LHC searches for SUSY and Higgs, a large fraction of SUSY parameter space is left unconstrained \cite{bs}. Therefore a search for the charged Higgs in the mass range presented in this paper is not sensitive to $B_s \rightarrow \mu^+\mu^-$ constraints.\\
The CP-odd Higgs boson mass is taken to be in the range $400~<~m_{A}~<~600$ GeV. The mass splitting between the neutral Higgs bosons in this case may arise large terms for Feynman diagrams involved in EW measurements such as the Z boson self-energy. We explicitely check the deviation from SM observations using the $\Delta \rho$ parameter. Fig. \ref{drho} shows results for $m_{H}=300$ GeV, \tanb = 10 and with three adopted values of $m_{A}$. The global fit to SM electroweak measurements requires $\Delta \rho$ to be $O(10^{-3})$ \cite{pdg}. Since Fig. \ref{drho} shows $\Delta \rho$ values within the same order of magnitude, therefore, we conclude that the set of chosen mass points are consistent with EW precision measurements.\\
As a summary, LHC (CMS and ATLAS) results based on MSSM Higgs boson searches, $b \ra s\gamma$ results from CLEO, $B_s \rightarrow \mu^+\mu^-$ measurements from LHCb, electroweak observations and $\Delta \rho$ measurements are all taken into account and a neutral Higgs boson with $m_{H}=300$ GeV and a charged Higgs boson in the mass range $170~ < ~m_{H^{\pm}}~<~400$ GeV is explored. The charged Higgs boson lighter than the CLEO lower bound (300 GeV) is included as this limit is not yet confirmed by direct searches for the charged Higgs at LHC. The paper is therefore dedicated to the search for a heavy charged and neutral Higgs boson not yet excluded by LHC experiments.\\
 
\section{Theoretical Framework}
Our theoretical basis is a two Higgs doublet model with the general potential as follows \cite{2hdm1,2hdm2}.
\begin{align}
\mathcal{V} = & m_{11}^2\Phi_{1}^{\dag}\Phi_{1} + m_{22}^2\Phi_{2}^{\dag}\Phi_{2} - \left[m_{12}^2\Phi_{1}^{\dag}\Phi_{2}+\textnormal{h.c.}\right] \notag \\
&+\frac{1}{2}\lambda_1\left(\Phi_{1}^{\dag}\Phi_{1}\right)^2+\frac{1}{2}\lambda_2\left(\Phi_{2}^{\dag}\Phi_{2}\right)^2
+\lambda_3\left(\Phi_{1}^{\dag}\Phi_{1}\right)\left(\Phi_{2}^{\dag}\Phi_{2}\right)
+\lambda_4\left(\Phi_{1}^{\dag}\Phi_{2}\right)\left(\Phi_{2}^{\dag}\Phi_{1}\right) \notag \\
&+\left\lbrace\frac{1}{2}\lambda_5\left(\Phi_{1}^{\dag}\Phi_{2}\right)^2+\left[\lambda_6\left(\Phi_{1}^{\dag}\Phi_{1}\right)
+\lambda_7\left(\Phi_{2}^{\dag}\Phi_{2}\right)\right]\left(\Phi_{1}^{\dag}\Phi_{2}\right)+\textnormal{h.c.}\right\rbrace 
\end{align} 
Hereafter, we use the usual abbreviations for parameters: $s_{\beta}\equiv \sin \beta$, $c_{\beta}\equiv \cos \beta$, $t_{\beta}\equiv \tan \beta$, $s_{\beta-\alpha}\equiv \sin(\beta-\alpha)$, $c_{\beta-\alpha}\equiv \cos(\beta-\alpha)$, $\lambda_{345}=\lambda_3+\lambda_4+\lambda_5$.
The free parameters of such a model are  
\be
\lambda_1,\lambda_2,\lambda_3,\lambda_4,\lambda_5,\lambda_6,\lambda_7,m_{12}^{2},\tb 
\ee
in the general basis. The CP violation or Flavor Changing Neutral Currents (FCNC) are not assumed as they are naturally suppressed via the Natural Flavor Conservation (NFC) mechanism by imposing a $Z_{2}$ symmetry on the Lagrangian \cite{weinberg,2hdm3} which leads to the following requirement,
\be
\lambda_6=\lambda_7=0.
\ee
By expressing $\lambda_i~(i \leq 5)$ in terms of the Higgs boson masses, $m_{12}^2$, $\alpha$, $\tb$, $\lambda_6$ and $\lambda_7$ \cite{2hdm4} the alternative set of free parameters can be taken as 
\be
m_h,m_H,m_A,m_{H^{\pm}},\alpha,\tb,m_{12}^2.
\ee
Further simplification can be made by observing that $m_{12}^2$ and $\lambda_5$ are related through the following relation:
\be
m_A^2=\frac{m_{12}^2}{\sinb \cosb}-\frac{v^2}{2}(2\lambda_5+\lambda_6 \tb^{-1}+\lambda_7\tb).
\ee
which reduces to $m_{12}^2=m_A^2 \cosb \sinb$ if $\lambda_5=\lambda_6=\lambda_7=0$. This requirement already exists in MSSM \cite{mm1,mm2,mm3} and has been used in triple Higgs studies \cite{HHH,HHH2}.

Therefore the following subset of parameters is enough to describe the model,
\be
m_h,m_H,m_A,m_{H^{\pm}},\alpha,\tb.
\ee
or equivalently 
\be
m_h,m_H,m_A,m_{H^{\pm}},\sbma,\tb.
\ee

\section{The Charged Higgs Production at $e^-e^+$ Linear Colliders}
The charged Higgs observation potential at a linear $e^{-}e^{+}$ collider such as the International Linear Collider ILC \cite{ilc,rdr} or the Compact Linear Collider (CLIC) \cite{clic} has been studied extensively in the literature. Most studies focus on charged Higgs pair production or in association with a gauge boson. A list of such studies can be found in \cite{pair,mine,sch6,sch7,sch1,sch2,sch3,sch4,sch5,sch8,WH1,WH2,WH3,WH4,WH5}. There are also studies of such processes at muon colliders in \cite{mine1,mine2,mine3}. The $H^+H^-$ pair produced in $e^+e^-$ collisions can be studied in two final states. The charged Higgs decay to $\tau \nu$ has a high branching ratio at low masses ($m_{H^{\pm}}<180$ GeV), however, it reduces to small values when heavy charged Higgs is going to be analyzed. This final state, i.e., $\tau^+ \tau^- E^{\textnormal{miss}}_{T}$ suffers from the large background including $W^+W^-$, $ZZ$ and Drell-Yan $Z^{(*)}/\gamma^*$ production. The charged Higgs decay to $t\bar{b}$ has a larger branching ratio for heavy charged Higgs cases. However, it produces high particle multiplicity, as there are a pair of top and bottom quarks in the final state.\\
In addition to studies of the charged Higgs bosons, double and triple neutral Higgs production has also been addressed in \cite{ref1,ref2}. Such processes contain gauge or neutrinos associated with a pair of neutral Higgs (double Higgs production) or a CP-odd associated with CP-even Higgs pairs (triple Higgs production). Study of such processes requires a separated analysis as their final states are different from what is considered here and can be done in the scope of a general 2HDM in a future work. Therefore the current analysis involves processes which contain charged Higgs bosons.  
\section{Triple Higgs Couplings} 
The triple Higgs production can be regarded as a unique process for the charged Higgs studies as in that case, it contains a pair of charged Higgs and a single neutral Higgs ($H^+H^-H^0$ or $H^+H^-h^0$). Having the neutral Higgs decayed to $b\bar{b}$, the final state has effectively two extra $b-$jets as compared to the charged Higgs pair production. This feature makes it easy to be distinguished from the background processes. The reason is lack of existence of true $b-$jets in SM background events $WW$, $ZZ$ and $Z^{(*)}/\gamma^*$. The $t\bar{t}$ background can be reduced by a cut on the invariant mass of the two $b$-jets.\\
The triple Higgs production has been studied in the context of linear colliders in different reports. In Ref. \cite{3H1}, radiative corrections to the triple Higgs coupling have been studied. The production cross section of triple Higgs production at $e^+e^-$ collisions has been studied in Ref. \cite{3H2,3H3}, where they analyze different sets of the Higgs boson masses and evaluate the cross section of different processes which involve three Higgs bosons as a function of the center of mass energy of a linear collider. The ratio of triple Higgs coupling in 2HDM to that in SM has been studied in detail in Ref. \cite{3H4} taking into account the perturbativity requirements on $\lambda_i$, vacuum stability and Higgs boson mass limits from direct and indirect searches. The effect of triple Higgs coupling in the production of Higgs pairs in 2HDM has been discussed in Ref. \cite{3H5} for different set of center of mass energies and integrated luminosities of a linear $e^+e^-$ collider. A similar study has also been performed in MSSM in Ref. \cite{3H6}. A different work, reports the effect of quantum corrections and triple Higgs self-interactions in the neutral Higgs pair production in 2HDM as a function of \tanb and $\lambda_5$ in Ref. \cite{3H7}.\\ 
In what follows, the triple Higgs production is analyzed as the main source of charged Higgs bosons. To this end, the triple Higgs couplings are used for the signal production as presented in \cite{3H8,3H9,3H10,3H2}. For a better comparison, we briefly present $H^0H^+H^-$ couplings presented in the mentioned references respectively in Eqns. \ref{1}, \ref{2}, \ref{3}, \ref{4}.

\begin{eqnarray}
g_{H^+H^-H^0}=\frac{2m_W s_W}{e}&[&s_{\beta-\alpha}(\frac{1}{4} s_{2\beta}^2 (\lambda_1+\lambda_2) + \lambda_{345} (s_{\beta}^4+c_{\beta}^4) - \lambda_4-\lambda_5 - s_{2\beta} c_{2\beta}(\lambda_6-\lambda_7)) \nonumber \\  
&+& c_{\beta-\alpha} (\frac{1}{2} s_{2\beta} (s_{\beta}^2\lambda_1-c_{\beta}^2\lambda_2+c_{2\beta}\lambda_{345})-\lambda_6 s_{\beta} s_{3\beta}-\lambda_7 c_{\beta} c_{3\beta}))] 
\label{1}
\end{eqnarray}

\be
g_{H^+H^-H^0}=\frac{m_W s_W s_{2\beta}}{e}[s_{\beta}c_{\alpha}\lambda_1 + c_{\beta} s_{\alpha} \lambda_2 -s_{\beta+\alpha}\lambda_{345} ] + c_{\beta-\alpha}\lambda_3
\label{2}
\ee

\be
g_{H^+H^-H^0}=\frac{e}{m_W s_W s_{2\beta}^2}(c_{\beta}^3 s_{2\beta}s_{\alpha}m_H^2+c_{\alpha}s_{2\beta}s_{\beta}^3m_{H}^2-2s_{\beta+\alpha}m_{12}^2+c_{\beta-\alpha}s_{2\beta}^2m_{H^{\pm}}^2)
\label{3}
\ee

\be
g_{H^+H^-H^0}=\frac{e}{m_{W} s_W s_{2\beta}}[(m_{H^{\pm}}^2-m_{A^{0}}^2+\frac{1}{2}m_{H^{0}}^2)s_{2\beta}c_{\beta-\alpha}-(m_{H^{0}}^2-m_{A^{0}}^2)c_{2\beta}s_{\beta-\alpha}]
\label{4}
\ee
The first two expressions are based on $\lambda_i$'s, while the second pair are written in terms of the Higgs boson masses. We check explicitly the above couplings and observe that they are equivalent when $\lambda_5=\lambda_6=\lambda_7=0$. Furthermore, the following relation is assumed:
\be
s_{\beta-\alpha}=1
\label{sba}
\ee
This equation makes sure that the neutral lightest Higgs boson has the same couplings to gauge bosons as the SM partner. It is therefore an SM-like Higgs boson. This is due to the fact that the ratio of Higgs-gauge coupling in 2HDM to SM Higgs-gauge coupling can be expressed as follows \cite{3H11}:\\
\be
\frac{g_{h_{2HDM}VV}}{g_{h_{SM}VV}}=s_{\beta-\alpha},~~\frac{g_{H_{2HDM}VV}}{g_{H_{SM}VV}}=c_{\beta-\alpha}
\ee
Since we use large $\beta$ values, the above requirement ($s_{\beta-\alpha}=1$) leads to small and negative $\alpha$ values. Moreover, in this limit (i.e., $s_{\beta-\alpha}=1$), the SM-like Higgs boson has the same coupling to a pair of bottom quarks as in SM, because \cite{3H11}\\
\be
\frac{g_{h_{2HDM}b\bar{b}}}{g_{h_{SM}b\bar{b}}}=-s_{\alpha}/c_{\beta}=s_{\beta-\alpha}-t_{\beta}c_{\beta-\alpha}
\ee
As all expressions quoted in Eqns. \ref{1}, \ref{2}, \ref{3} and \ref{4} are equivalent in the region of parameter space studied in this paper (i.e., $\lambda_5=\lambda_6=\lambda_7=0$ and $s_{\beta-\alpha}=1$), Eq. \ref{4} is used for computational purposes. 
\section{Event Simulation}
The triple Higgs coupling presented in Eq. \ref{4} is implemented in CompHEP 4.5.1 \cite{comphep,comphep2} for signal event generation at the hard interaction stage and cross section calculation. The double Higgs production ($H^{\pm}W^{\mp}H^{0}$) is also simulated by CompHEP. The output of CompHEP in LHEF format \cite{lhef} is used by PYTHIA 8.1.53 \cite{pythia} for further decay processing and particle showering, initial and final state radiation and multiple interactions. The background events are, however, simulated solely by PYTHIA which is used for both event generation and their cross section calculation. The jet reconstruction is performed using FASTJET 2.4.1 \cite{fastjet} with the anti-kt algorithm \cite{antikt}, a cone size of 0.4, and $E_T$ recombination scheme. For the calculation of the particle spectrum, the renormalization group evolution program SuSpect \cite{suspect} is used. The output including the particles mass spectra and decays is written in SLHA format \cite{slha} and used by PYTHIA for event generation. The neutral and charged Higgs branching ratio of decays are calculated by 2HDMC 1.1 \cite{2hdmc}. \\

\section{Signal and Background Events and their Cross Sections}
\begin{figure}[h]
 \begin{center}
 \includegraphics[width=0.6\textwidth]{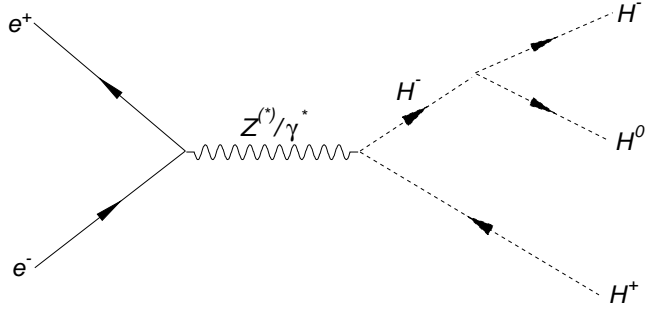}
 \end{center}
 \caption{The $e^+ e^- \ra H^+H^-H^0$ production process.}
 \label{diagram}
 \end{figure}

\begin{figure}[h]
 \begin{center}
 \includegraphics[width=0.6\textwidth]{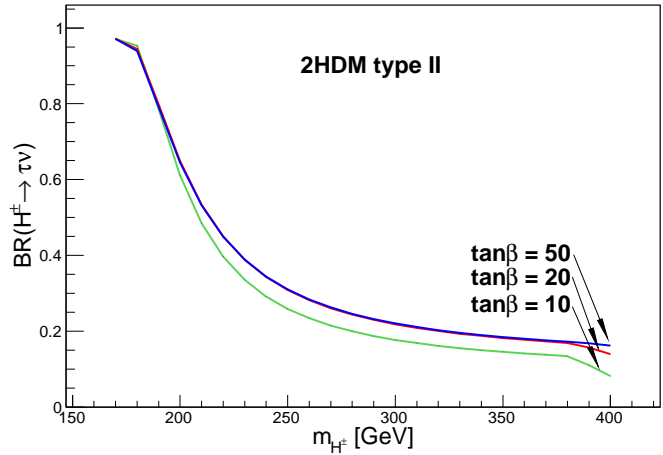}
 \end{center}
 \caption{The charged Higgs branching ratio of decay to $\tau \nu$.}
 \label{br}
 \end{figure}

\begin{figure}[h]
 \begin{center}
 \includegraphics[width=0.6\textwidth]{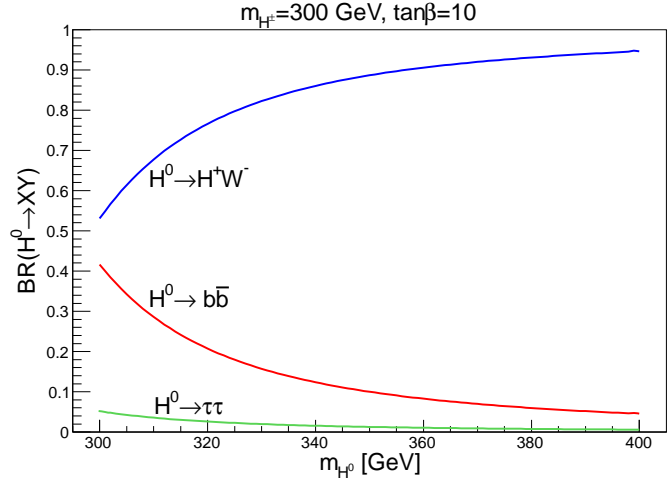}
 \end{center}
 \caption{The neutral Higgs branching ratio of decays.}
 \label{brhbb}
 \end{figure}

\begin{figure}[h]
 \begin{center}
 \includegraphics[width=0.6\textwidth]{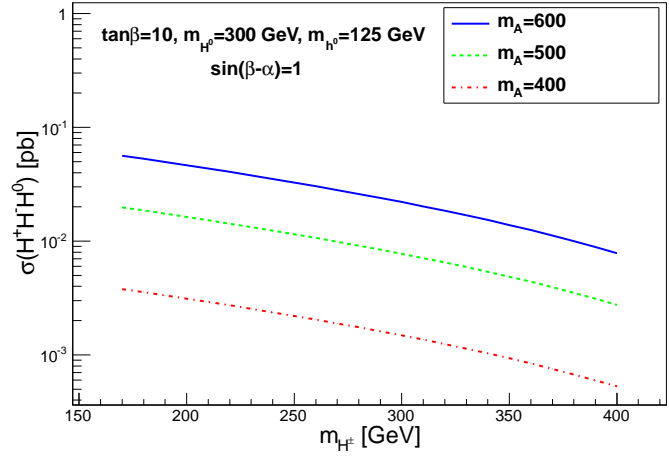}
 \end{center}
 \caption{The signal cross section with \tanb=10}
 \label{s1}
 \end{figure}

\begin{figure}[h]
 \begin{center}
 \includegraphics[width=0.6\textwidth]{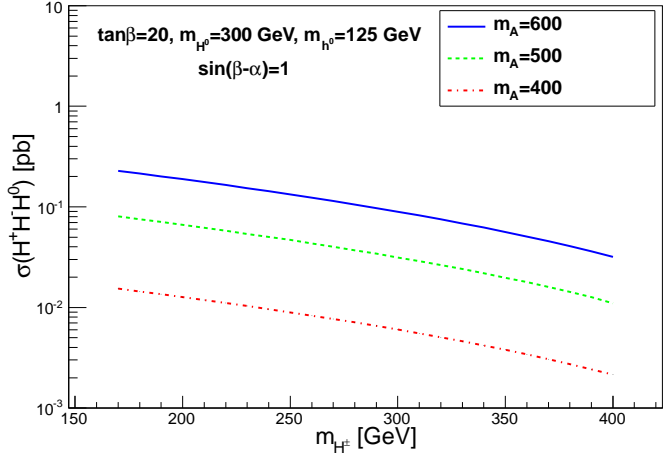}
 \end{center}
 \caption{The signal cross section with \tanb=20}
 \label{s2}
 \end{figure}

\begin{figure}[h]
 \begin{center}
 \includegraphics[width=0.6\textwidth]{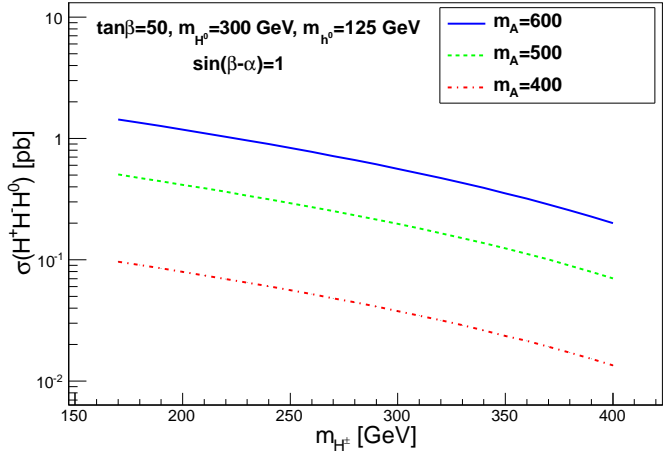}
 \end{center}
 \caption{The signal cross section with \tanb=50}
 \label{s3}
 \end{figure}

\begin{figure}[h]
 \begin{center}
 \includegraphics[width=0.6\textwidth]{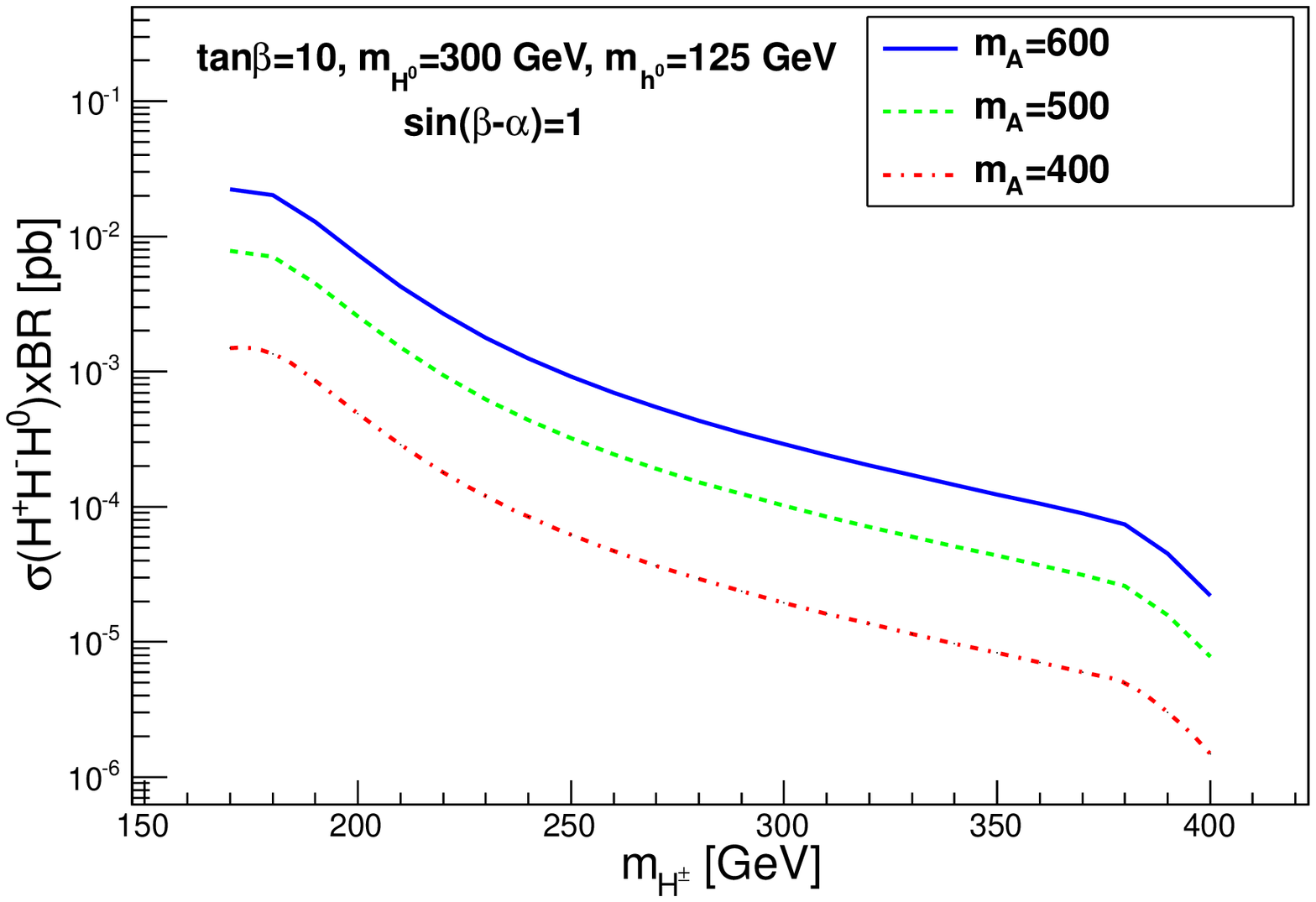}
 \end{center}
 \caption{The signal cross section times branching ratio of charged and neutral Higgs decays with \tanb=10}
 \label{sb1}
 \end{figure}

\begin{figure}[h]
 \begin{center}
 \includegraphics[width=0.6\textwidth]{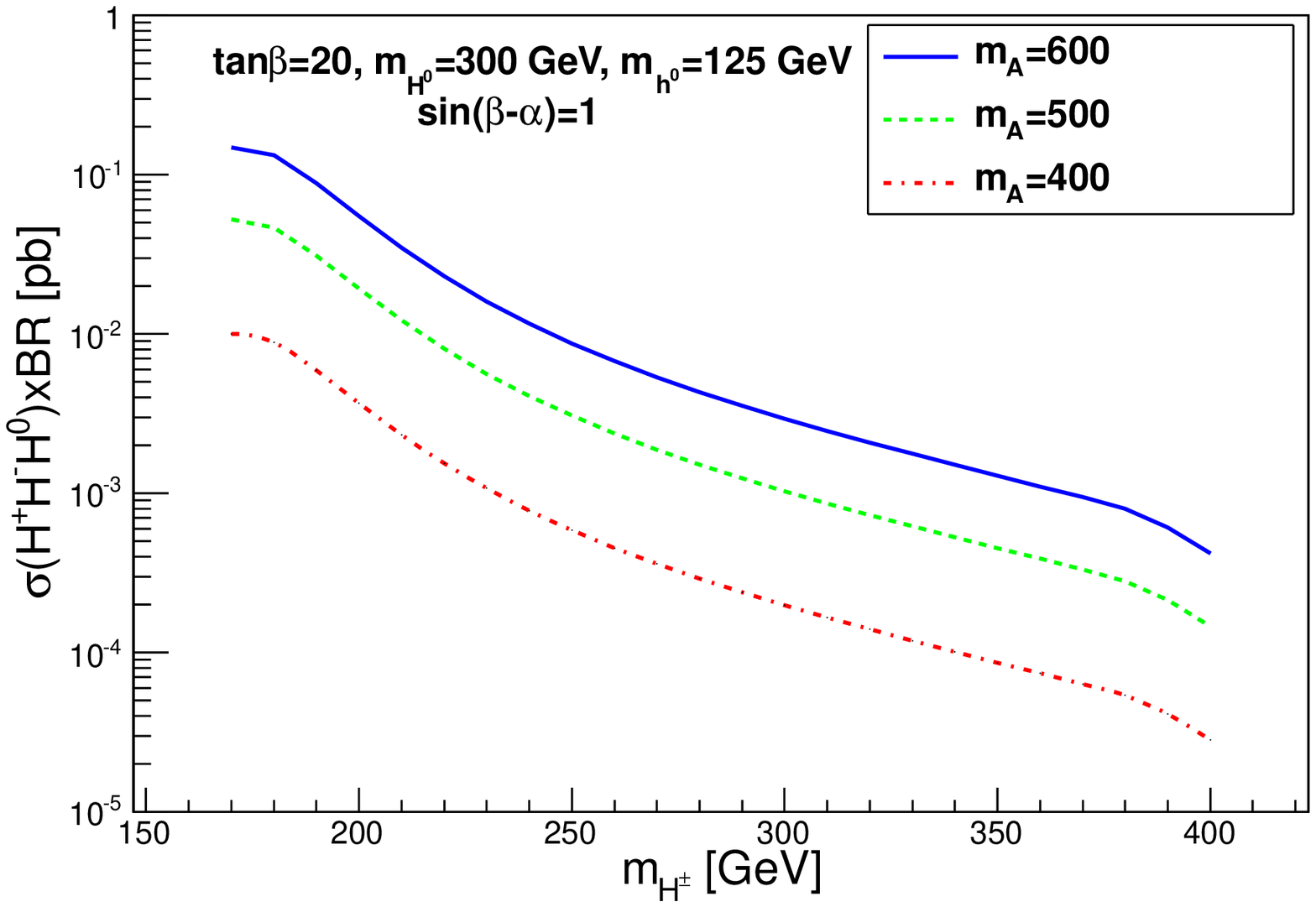}
 \end{center}
 \caption{The signal cross section times branching ratio of charged and neutral Higgs decays with \tanb=20}
 \label{sb2}
 \end{figure}

\begin{figure}[h]
 \begin{center}
 \includegraphics[width=0.6\textwidth]{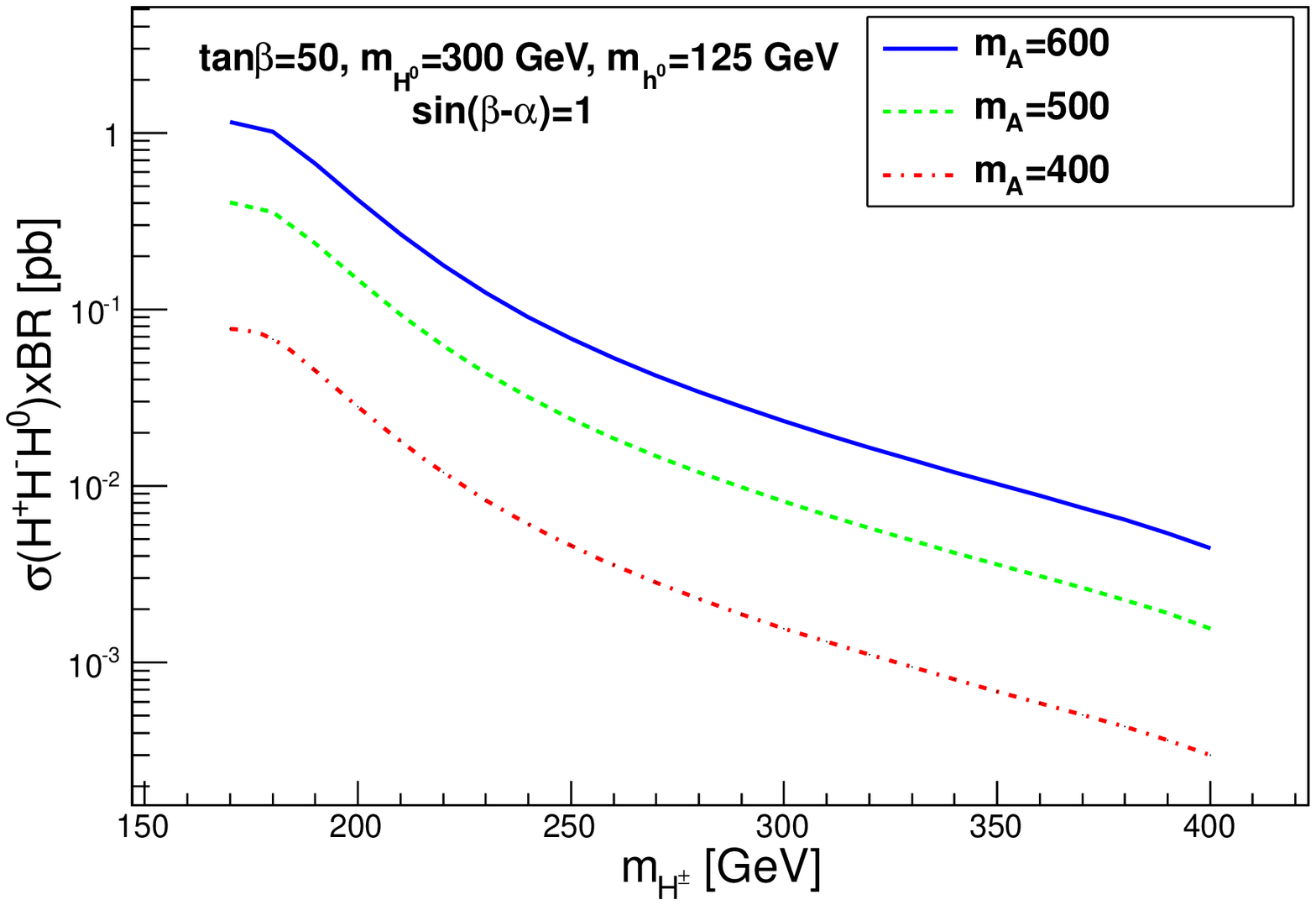}
 \end{center}
 \caption{The signal cross section times branching ratio of charged and neutral Higgs decays with \tanb=50}
 \label{sb3}
 \end{figure}

\begin{figure}[h]
 \begin{center}
 \includegraphics[width=0.6\textwidth]{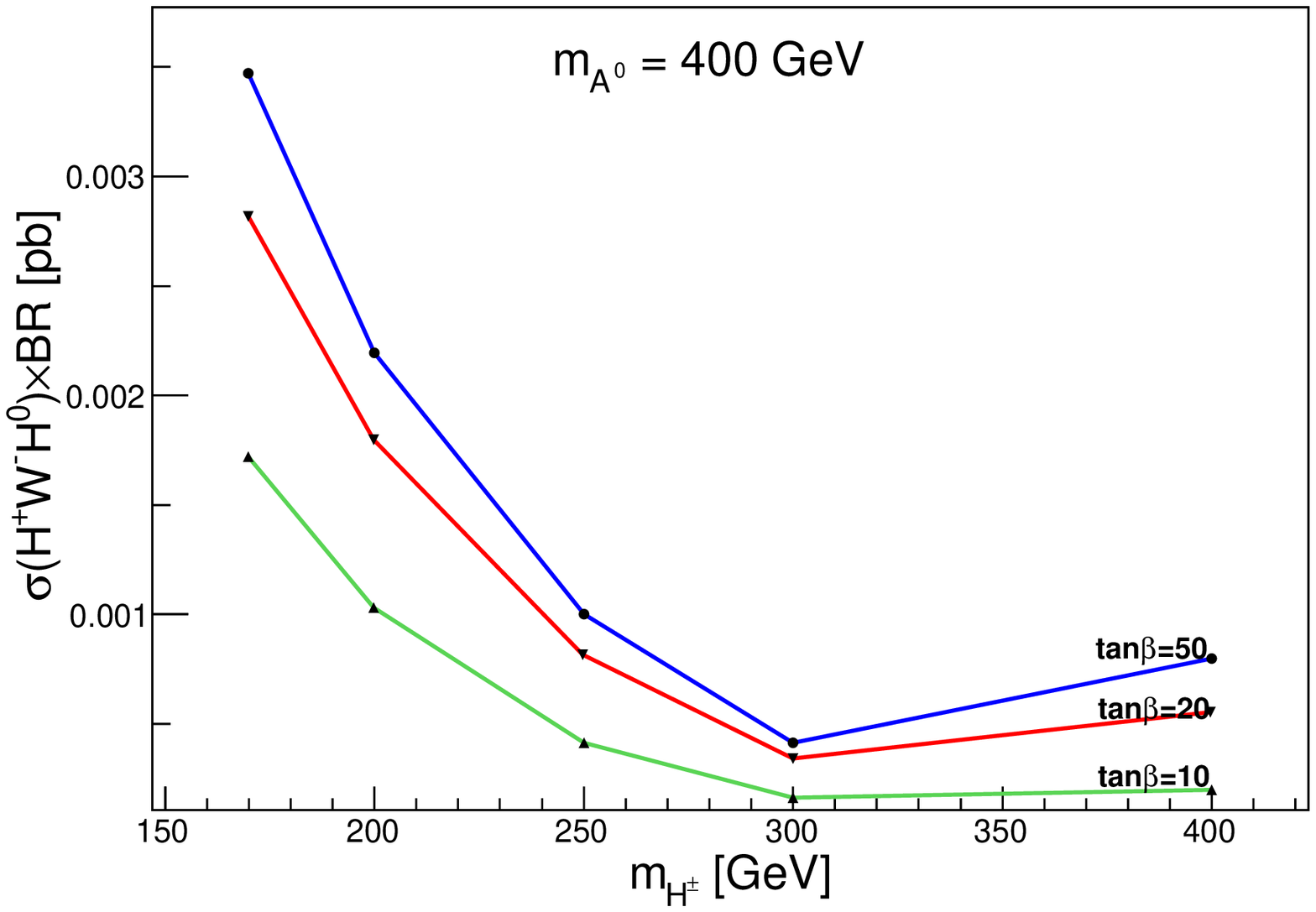}
 \end{center}
 \caption{The $H^+W^-H^0$ signal cross section times branching ratio of charged and neutral Higgs decays.}
 \label{HWH}
 \end{figure}

The triple Higgs production can be either $H^+H^-H^0$ or $H^+H^-h^0$. However, according to the corresponding couplings presented in Eqns. \ref{HHH} and \ref{HHh} (Ref. \cite{3H2}), the $H^+H^-H^0$ coupling is larger than $H^+H^-h^0$ in the limit $s_{\beta-\alpha}=1$, unless there is a very large mass difference between the charged Higgs and CP-odd neutral Higgs.
\be
H^{\pm}H^{\pm}H^0:~\frac{-ie}{m_W \sw s_{2\beta}}\left[(m_{H^{\pm}}^2-m_{A}^2+\frac{1}{2}m_{H}^2)s_{2\beta}\cbma-(m_{H}^2-m_{A}^2)c_{2\beta}\sbma \right] 
\label{HHH}
\ee 
\be
H^{\pm}H^{\pm}h^0:~\frac{-ie}{m_W \sw s_{2\beta}}\left[(m_{H^{\pm}}^2-m_{A}^2+\frac{1}{2}m_{h}^2)s_{2\beta} \sbma +(m_{h}^2-m_{A}^2) c_{2\beta}\cbma \right] 
\label{HHh}
\ee 
Therefore the signal is assumed to be $H^+H^-H^0$ or $H^+W^-H^0$ production. The latter process involves one charged Higgs but its cross section is comparable to the triple Higgs process. Therefore by signal we mean a sum of the above two processes. In order to have a reasonable background rejection, it is a convenient choice to assume the neutral Higgs decay to $b\bar{b}$. Since the charged Higgs decay to $t\bar{b}$ produces a high multiplicity event, it is better to choose $H^{\pm}\ra \tau \nu$ decays which produce $\tau$-jets and $E^{\textnormal{miss}}_{T}$ in the final state. According to the Higgs-fermion Yukawa couplings defined for the four types of 2HDM as in Tab. \ref{types} (Ref. \cite{main2hdm}), the type-II 2HDM is most suitable for such a final state, because it provides the largest $H^{\pm} \ra \tau\nu$ and $H^0 \ra b \bar{b}$ branching ratio of decays at high \tanb. 
\begin{table}[h]
\begin{tabular}{|c|c|c|c|c|}
\hline
\multicolumn{5}{|c|}{Type}\\
& I & II & III & IV \\
\hline
$\rho^D$ & $\kappa^D \ctb$ & $-\kappa^D \tb$ & $-\kappa^D \tb$ & $\kappa^D \ctb$ \\
\hline  
$\rho^U$ & $\kappa^U \ctb$ & $\kappa^U \ctb$ & $\kappa^U \ctb$ & $\kappa^U \ctb$ \\
\hline  
$\rho^L$ & $\kappa^L \ctb$ & $-\kappa^L \tb$ & $\kappa^L \ctb$ & $-\kappa^L \tb$ \\
\hline  
\end{tabular}
\caption{The four types of a general 2HDM in terms of the couplings in the Higgs-fermion Yukawa sector. \label{types}}
\end{table}
\\ As a summary the full signal production process is  
\be
e^+e^- \ra H^+H^-H^0 (H^+W^-H^0) \ra \tau^+ \tau^- b \bar{b} E^{\textnormal{miss}}_{T}
\ee
The Feynman diagram of signal events (triple Higgs production) is shown in Fig. \ref{diagram}. The Higgs coupling depends on $m_A^2 - m_H^2$ according to Eq. \ref{4}. The neutral Higgs mass ($m_{H^0}$) has to be greater than $m_{h^0}$, and be small enough to allow for heavy charged Higgs masses to be produced at the linear collider, however, it is constrained from below by LHC searches for neutral $H \ra \tau\tau$. A neutral Higgs with $m_{H^0}=300$ GeV is well outside the excluded area. Therefore, we set $m_{H^0}=300$ GeV throughout the paper. Now the cross section can increase if $m_A$ increases resulting in larger $m_A^2 - m_H^2$ factors. Therefore the signal total cross section has been plotted as a function of the charged Higgs mass for different \tanb and $m_A$ values in Figs. \ref{s1}, \ref{s2}, \ref{s3}. The charged Higgs branching ratio of decay to $\tau\nu$ at 2HDM type-II is shown in Fig. \ref{br}. Other decay channels such as $H^{+} \ra W^+H^0$ and $H^+ \ra W^+A^0$ lead to three or four particles for each charged Higgs decay and are not considered as their identification is difficult due to limited particle identification efficiencies and detector considerations. The main decay channels of the neutral Higgs are also shown in terms of branching ratios in Fig. \ref{brhbb}. Similar results are observed for other values of \tanb. The neutral Higgs branching ratio of decay to $b\bar{b}$ decreases with increasing Higgs boson mass, therefore higher cross sections are expected for lighter neutral Higgs bosons. The decay channel $H \ra H^{\pm}W^{\mp} $ leads to three charged Higgs bosons in the final state of the main process and is not considered here although it acquires a higher branchig ratio than $b\bar{b}$. Other decay channels, e.g., $H^{0} \ra h^{0}h^{0}$, $H^{0} \ra W^{+}W^{-}$ and $H^{0} \ra Z^{0}Z^{0}$ vanish as they are proportional to $c_{\beta-\alpha}$ and this analysis is based on $s_{\beta-\alpha}=1$. Therefore as long as the lightest neutral Higgs is required to be SM-like with $s_{\beta-\alpha}=1$, such decay channels do not play a role. The cross section times branching ratio of the triple Higgs production is thus obtained using branching ratios presented in Figs. \ref{br},\ref{brhbb}. Results are shown in Figs. \ref{sb1}, \ref{sb2} and \ref{sb3}. The double Higgs production originates from different decay chains, $e^+e^- \ra A^0 H^0 \ra H^+ W^- H^0$ and $e^+e^- \ra H^+ H^- \ra H^+ W^- H^0$. Therefore, its cross section depends on $m_{A^{0}}$, $m_{H^{0}}$ and $m_{H^{\pm}}$. Figure \ref{HWH} shows the cross section of this process times BR$(H^{\pm}\ra \tau \nu)\times$BR$(H^{0}\ra b\bar{b})$. A double charged Higgs production through $e^+e^- \ra H^+ H^- \ra t \bar{b} \tau \bar{\nu} \ra W^+ b \bar{b} \tau \bar{\nu} \ra \tau^+ \nu b \bar{b} \tau \bar{\nu}$ was also explicitly checked and turned out to make no contribution to the signal as it was suppressed by the cut on the $b\bar{b}$ invariant mass.   


The background events are SM processes, $ZZ$, $WW$, $Z^{(*)}/\gamma^*$ and $t\bar{t}$ with cross sections 0.13, 1.8, 2.0 and 0.1  $pb$ respectively at $\sqrt{s}=1.5$ TeV. If $Z\ra \tau \tau$ and $Z \ra b\bar{b}$ decays occur, the $ZZ$ process can lead to $\tau \tau b \bar{b}$ final state. The $t\bar{t}$ events are also important background events as they contain two $b$-jets. Other sources of triple Higgs, e.g., $h^{0}H^+H^-$ and $A^{0}H^+H^-$ turn out to be negligible with a cross section of the order $10^{-6}$ pb. 
\section{Event Selection and Analysis}
\begin{figure}[h]
 \begin{center}
 \includegraphics[width=0.6\textwidth]{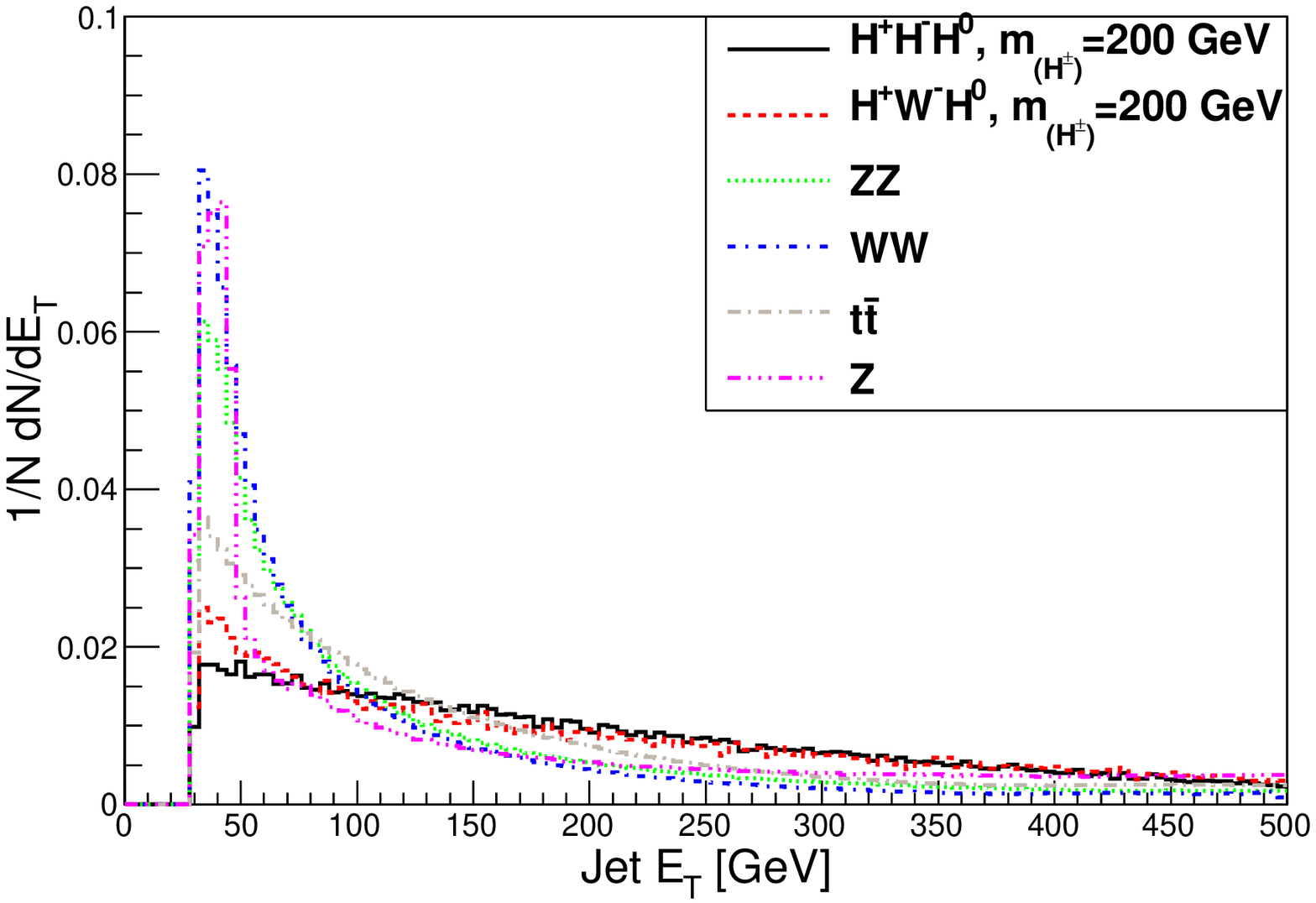}
 \end{center}
 \caption{The jet transverse energy distribution in signal and background events.}
 \label{jetet}
 \end{figure}

\begin{figure}[h]
 \begin{center}
 \includegraphics[width=0.6\textwidth]{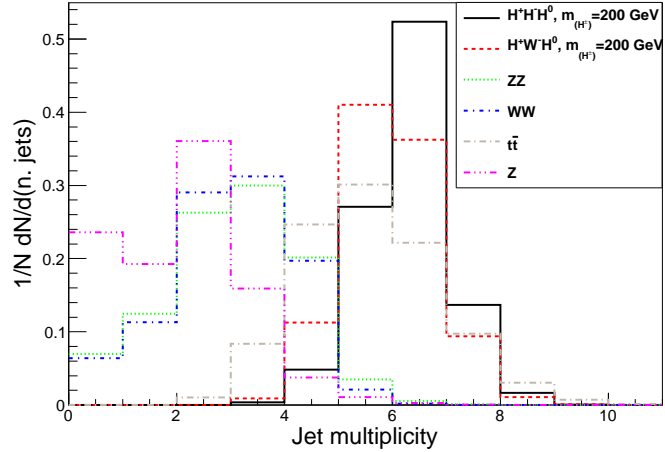}
 \end{center}
 \caption{The jet multiplicity in signal and background events.}
 \label{jetmul}
 \end{figure}

\begin{figure}[h]
 \begin{center}
 \includegraphics[width=0.6\textwidth]{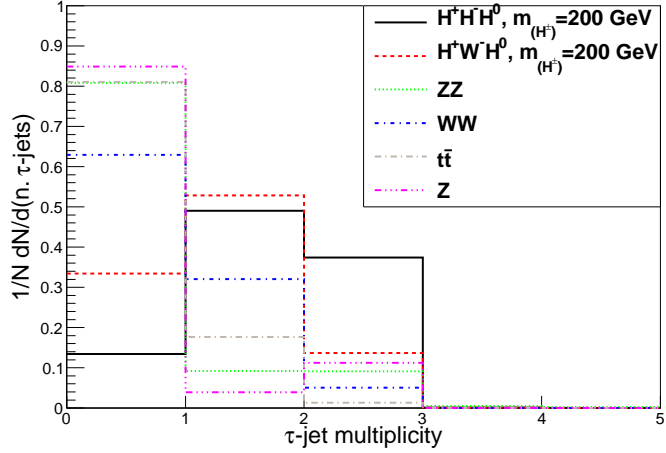}
 \end{center}
 \caption{The $\tau$-jet multiplicity in signal and background events.}
 \label{taujetmul}
 \end{figure}

\begin{figure}[h]
 \begin{center}
 \includegraphics[width=0.6\textwidth]{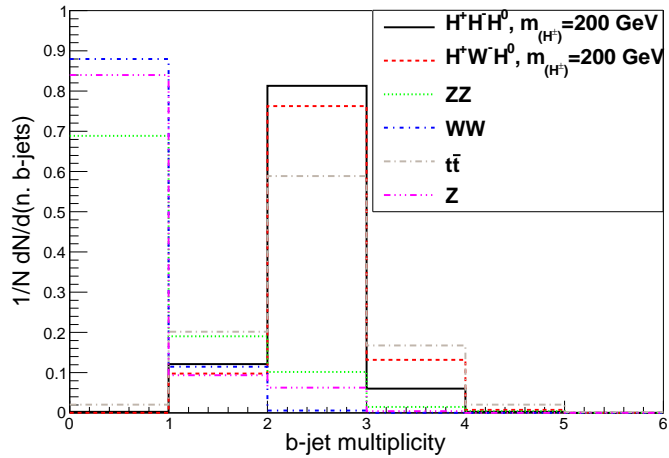}
 \end{center}
 \caption{The $b$-jet multiplicity in signal and background events.}
 \label{bjetmul}
 \end{figure}

\begin{figure}[h]
 \begin{center}
 \includegraphics[width=0.6\textwidth]{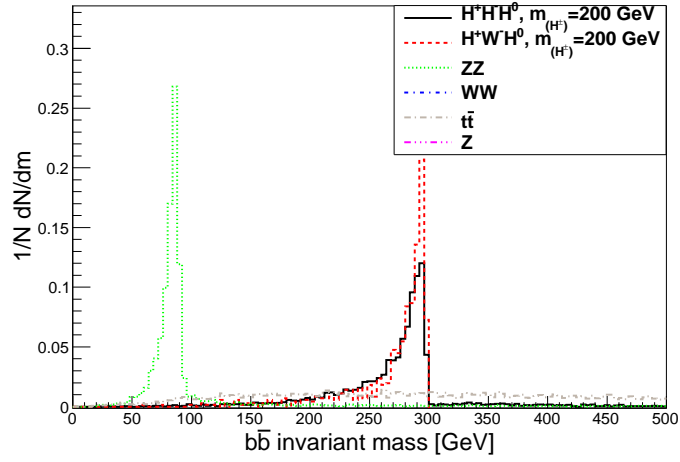}
 \end{center}
 \caption{The $b$-jet pair invariant mass distribution in signal and background events.}
 \label{bbm}
 \end{figure}

\begin{figure}[h]
 \begin{center}
 \includegraphics[width=0.6\textwidth]{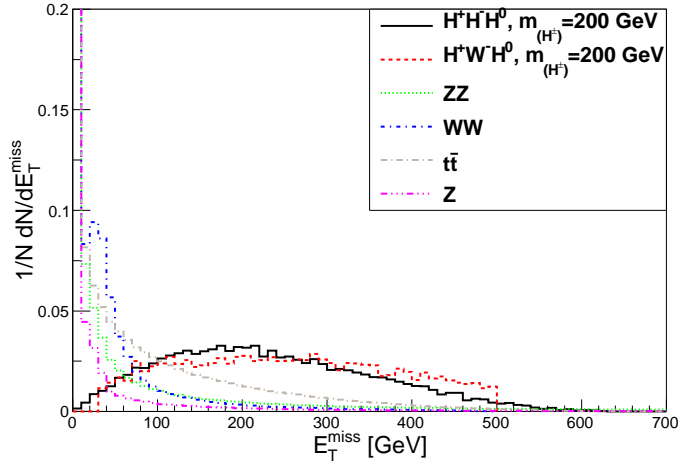}
 \end{center}
 \caption{The missing transverse energy distribution in signal and background events.}
 \label{met}
 \end{figure}

Signal events contain four jets: two $\tau$-jets and two $b$-jets. Therefore three requirements on the number of jets are to be applied to separate signal and background, i.e., the cut on the number of all jets (with a kinematic cut as deduced from the jet transverse energy distributions), the cut on the number of $\tau$-jets which are signatures of the charged Higgs and a separated cut on the number of $b$-jets. The two $b$-jets originate from a neutral heavy Higgs boson in signal events, therefore, their invariant mass should lie within a mass window tuned by the neutral Higgs mass. Finally as there are two neutrinos in the event, a requirement on the minimum missing transverse energy should help suppression of some background events like single or double $Z$ bosons. From detector point of view, such requirements imply experimental uncertainties due to the jet energy scale, $b$-tagging efficiency, $\tau$-identification efficiency and missing transverse energy resolution. These uncertainties should be taken into account in a real analysis when a reasonable knowledge of their values is achieved. \\
In order to start event selection kinematic distributions are studied. Figure \ref{jetet} shows the (any) jet transverse energy distribution. Therefore the first step in signal selection is to require at least four jets in the final state with kinematic cuts on the jet transverse energy and pseudorapidity as in Eq. \ref{jetetcut}. The pseudorapidity is defined as $\eta=-\ln \tan (\theta/2)$ and $\theta$ is the polar angle evaluated with the respect to the beam axis. This requirement is basically applied in order to reject soft and forward/backward jets whose reconstruction error may be large. In addition, there are SM backgrounds which are suppressed by the kinematic cuts on jets, as they produce softer jets than those of the signal events. As seen from Fig. \ref{jetet}, background events tend to produce jets with $E^{\textnormal{jet}}_{T}\simeq 50$ GeV because the decaying particle, which produces the jet, is either $Z$ or $W$ boson. A harder cut may suppress a large fraction of such backgrounds, however, the signal cross section is small at heavy charged Higgs area. Therefore we avoid hard cuts in order to keep the signal events at a reasonable minimum.

\be
E^{\textnormal{jet}}_{T}~>~30 \textnormal{GeV},~~~~|\eta|<3
\label{jetetcut}
\ee 
Selected jets are counted in the second step. Figure \ref{jetmul} shows the number of reconstructed jets passing the requirement of Eq. \ref{jetetcut}. A cut on the number of reconstructed jets is applied as in Eq. \ref{jetmulcut}.
\be
\textnormal{Number of jets (satisfying Eq. \ref{jetet}}) \geq 4
\label{jetmulcut}
\ee
In the next step, a $\tau$-ID algorithm is applied to jets, similar to the algorithm used by LHC experiments \cite{tauid}. The algorithm starts with a cut on the transverse energy of the hardest charged particle track in the $\tau$-jet cone as $E_{T} > 20 \textnormal{GeV}$. This requirement is basically applied as we expect a low charged particle multiplicity in $\tau$ hadronic decay which results in a large fraction of the $\tau$ energy to be carried by the leading track (the charged pion). The isolation requirement further uses the above feature of the $\tau$ hadronic decay by requiring no track with $p_T>1$ GeV to be in the annulus defined as $0.1< \Delta R <0.4$. Here $\Delta R=\sqrt{(\Delta \eta)^2 + (\Delta \phi)^2}$ and $\phi$ is the azimuthal angle. The $\Delta$ is calculated between the cone surface and the cone axis defined by the hardest track. The number of signal tracks are then calculated by searching for tracks in the cone defined around the hardest track with $\Delta R <0.07$. Since $\tau$ leptons decay predominantly to one or three charged pions, we require the number of signal tracks to be one or three. A jet (a $\tau$ lepton candidate) has to pass all above requirements to be selected as a $\tau$ lepton. Figure \ref{taujetmul} shows the $\tau$-jet multiplicity in signal and background events. Finally we require that there should be two $\tau$-jets satisfying all above requirements in the event.
The selection of $b$-jets is started at this stage, by selecting jets which are within $\Delta R < 0.4$ with respect to the generated $b$ or $c$ quarks. A jet is tagged as a $b$-jet with a probability of 60$\%$ (10$\%$) if it matches a $b$-quark ($c$-quark). The above numbers are assumed as the $b$-tagging efficiency and fake rate respectively. Figure \ref{bjetmul} shows the number of $b$-tagged jets in signal and background events. For the signal selection, we require two $b$-jets in the event.\\
In the next step the above two jets are used for $b\bar{b}$ invariant mass reconstruction. Figure \ref{bbm} shows the distribution of the $b$-jet pair invariant mass in signal and the only remaining background at this stage. As seen from Fig. \ref{bbm}, the $b$-jet pair invariant mass tends to peak at 150 GeV (the neutral Higgs boson mass) in signal events while for the $ZZ$ background the $b\bar{b}$ invariant mass obviously peaks at the $Z$ mass. Based on this observation the requirement presented in Eq. \ref{bbcut} is applied on the distribution of $b$-jet pair invariant mass.  
\be
b\bar{b} ~\textnormal{invariant mass} > 120 ~\textnormal{GeV}
\label{bbcut}
\ee
As the last step, the missing transverse energy is reconstructed as the negative vectorial sum of particle momenta in the transverse plane as shown in Fig. \ref{met}.
Based on the distribution shown in Fig. \ref{met}, the requirement of Eq. \ref{metcut} is applied on signal and background events.
\be
E^{\textnormal{miss}}_T~>~30~ \textnormal{GeV}
\label{metcut}
\ee
\section{Results}
\begin{figure}[h]
 \begin{center}
 \includegraphics[width=0.6\textwidth]{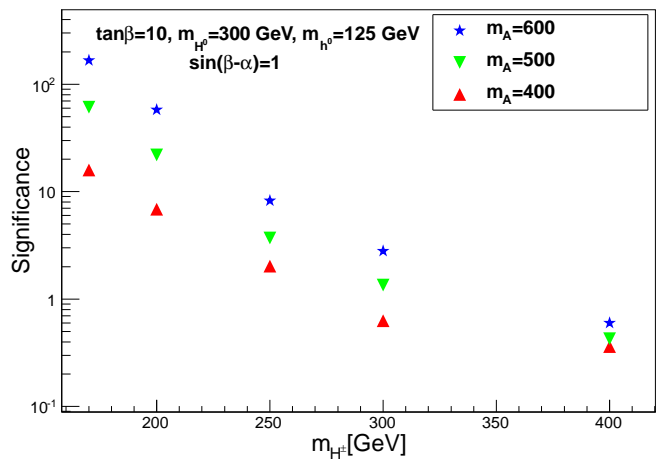}
 \end{center}
 \caption{The signal significance with \tanb=10, as a function of the charged Higgs mass and with different $m_A$ values at $\sqrt{s}=1.5$ TeV and integrated luminosity 500 $fb^{-1}$.}
 \label{sig1}
 \end{figure}

\begin{figure}[h]
 \begin{center}
 \includegraphics[width=0.6\textwidth]{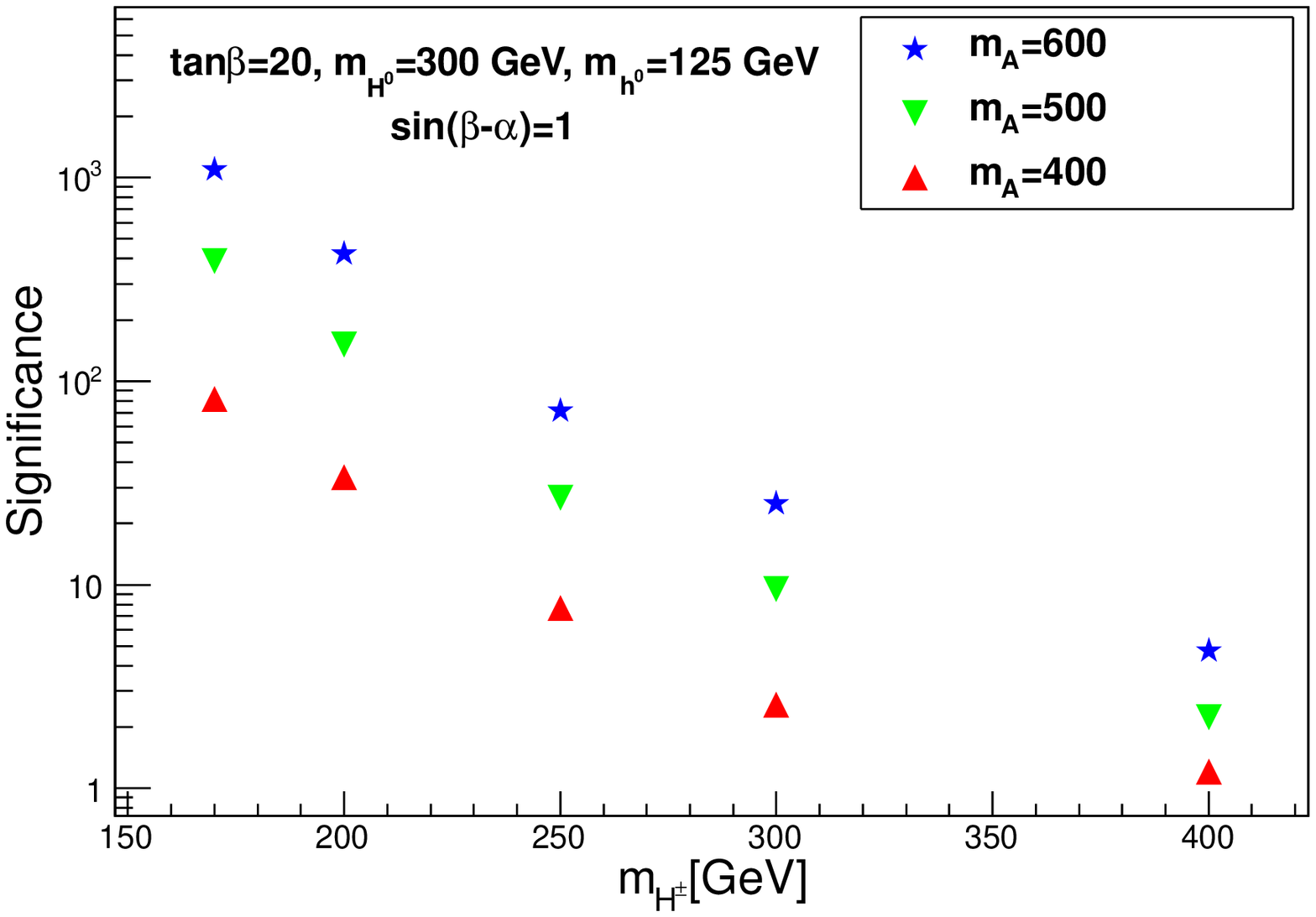}
 \end{center}
 \caption{The signal significance with \tanb=20, as a function of the charged Higgs mass and with different $m_A$ values at $\sqrt{s}=1.5$ TeV and integrated luminosity 500 $fb^{-1}$.}
 \label{sig2}
 \end{figure}

\begin{figure}[h]
 \begin{center}
 \includegraphics[width=0.6\textwidth]{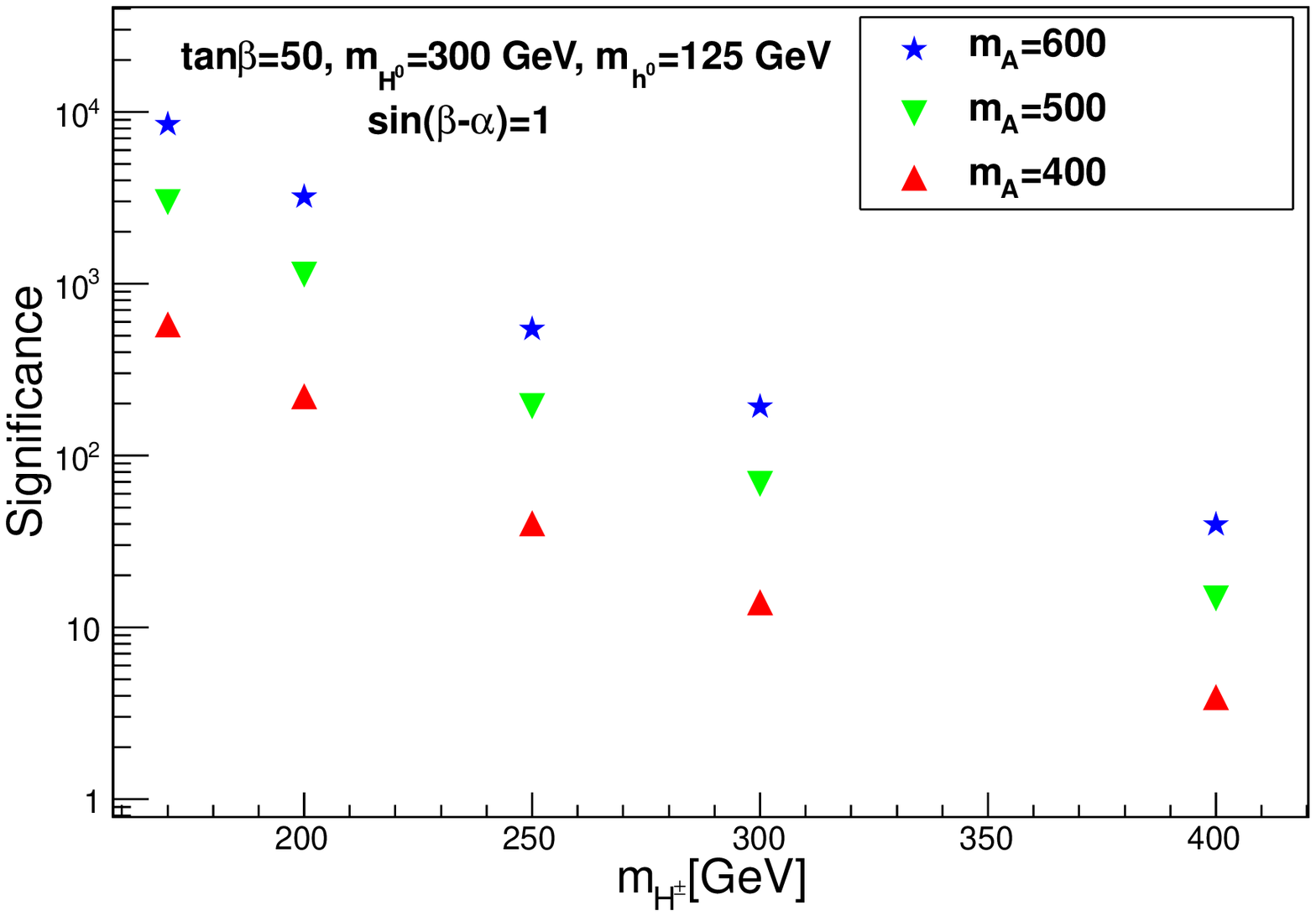}
 \end{center}
 \caption{The signal significance with \tanb=50, as a function of the charged Higgs mass and with different $m_A$ values at $\sqrt{s}=1.5$ TeV and integrated luminosity 500 $fb^{-1}$.}
 \label{sig3}
 \end{figure}

An event has to pass all requirements in the previous section to be selected. Selection cuts are applied one after the other, and relative efficiencies and the total efficiency of the signal and background selection is calculated. The final number of events, of course depends not only on the total selection efficiency, but also on the cross section of events. In case of signal, the cross section depends on \tanb, $m_{H^{\pm}}$ and $m_A$ and branching ratio of Higgs decays. Table \ref{seleff} shows the signal and background selection efficiencies.
\begin{table}[h]
\begin{tabular}{|c|c|c|c|c|c|c|c|c|c|c|c|c|c|c|}
\hline
& \multicolumn{10}{c|}{Signal, $m_{H^{\pm}}:$} & \multicolumn{4}{c|}{} \\
& \multicolumn{2}{c|}{170 GeV} & \multicolumn{2}{c|}{200 GeV} & \multicolumn{2}{c|}{250 GeV} & \multicolumn{2}{c|}{300 GeV} & \multicolumn{2}{c|}{400 GeV} & \multicolumn{4}{c|}{Background} \\
\hline
&  HHH & HHW & HHH & HHW & HHH & HHW & HHH & HHW & HHH & HHW & ZZ & $Z^{(*)}/\gamma^*$ & WW & $t\bar{t}$ \\
\hline
Four jets & 0.64 & 0.64 & 0.63 & 0.63 & 0.63 & 0.59 & 0.63 & 0.63 & 0.62 & 0.62 & 0.24 & 0.052 & 0.22 & 0.91 \\
\hline
Leading track & 0.99 & 1 & 0.99 & 1 & 0.99 & 1 & 0.99 & 1 & 0.99 & 0.99 & 0.87 & 0.95 & 0.92 & 0.96 \\
\hline
Isolation & 0.87 & 0.69 & 0.88 & 0.69 & 0.89 & 0.78 & 0.9 & 0.63 & 0.9 & 0.83 & 0.35 & 0.092 & 0.6 & 0.24 \\
\hline
Number of signal tracks& 0.99 & 0.97 & 0.99 & 0.97 & 0.99 & 0.97 & 0.99 & 0.96 & 0.99 & 0.99 & 0.87 & 0.37 & 0.95 & 0.87 \\
\hline
Two $\tau$-jets & 0.41 & 0.2 & 0.43 & 0.21 & 0.45 & 0.28 & 0.47 & 0.22 & 0.5 & 0.1 & 0.68 & 0.6 & 0.19 & 0.071 \\
\hline
$\tau$-jet charge & 1 & 0.99 & 1 & 0.99 & 1 & 1 & 1 & 0.99 & 1 & 0.99 & 1 & 1 & 1 & 0.97 \\
\hline
Two $b$-jets & 0.82 & 0.84 & 0.82 & 0.84 & 0.82 & 0.85 & 0.82 & 0.82 & 0.82 & 0.81 & 0.15 & 0 & 0 & 0.61 \\
\hline
$b\bar{b}$ inv. mass & 0.88 & 0.94 & 0.88 & 0.93 & 0.88 & 0.94 & 0.87 & 0.93 & 0.85 & 0.96 & 0.041 & nan & nan & 0.33 \\
\hline
$E^{\textnormal{miss}}_{T}$ & 0.99 & 0.98 & 0.98 & 0.99 & 0.98 & 0.83 & 0.99 & 1 & 0.98 & 0.99 & 0.28 & nan & nan & 0.9 \\
\hline
Total eff. & 0.16 & 0.064 & 0.17 & 0.067 & 0.18 & 0.082 & 0.18 & 0.063 & 0.19 & 0.039 & 7.7e-05 & 0 & 0 & 0.0023 \\
\hline
\end{tabular}
\caption{Signal and background selection efficiencies. HHH and HHW mean triple and double Higgs processes respectively. The signal selection efficiencies are assumed to be independent of \tanb and $m_A$. \label{seleff}}
\end{table}
The selection efficiencies are used in the next step to calculate the number of signal and background events at a given point in parameter space. The signal significance is calculated as $N_S/\sqrt{N_B}$, where $N_S(N_B)$ is the signal (background) number of events after all selection cuts. The significance depends on \tanb and $m_A$ due to the dependence of cross section to these parameters. Therefore different plots are produced for each value of \tanb and $m_A$ as shown in Figs. \ref{sig1}, \ref{sig2}, \ref{sig3}.
\section{Conclusion}
The triple Higgs boson production was analyzed as a source of charged Higgs pairs. The analysis was performed for a linear $e^+e^-$ collider operating at $\sqrt{s}=1.5$ TeV and results were presented with a normalization to an integrated luminosity of 500 $fb^{-1}$.\\
The theoretical framework was set to 2HDM type-II containing an SM-like light Higgs boson with a mass equal to the current LHC observations. The effect of the CP-odd neutral Higgs mass in the production cross section and the signal significance was studied and it was concluded that increasing $m_A$ could increase the signal significance very sizably. The signal significance depends also on \tanb.\\
Thanks to the $b$-tagging tools, a reasonable background suppression is achieved leading to high signal significance values for some areas of the parameter space which correspond to heavy CP-odd neutral Higgs and high \tanb. Finally the signal significance was presented as a function of the charged Higgs mass, $m_A$ and \tanb. \\
The analysis reveals the future linear colliders potential for a heavy charged Higgs observation if LHC fails to do so. If the charged Higgs is heavier than 300 GeV, it may escape from LHC experiments. In such a scenario a linear collider with enough integrated luminosity higher than 500 $fb^{-1}$ would probably be the only experiment which could provide some news about this particle in the future.

\end{document}